\documentclass[preprint]{aastex}
\newcommand{\snr}{SNR 0540-69.3}
\newcommand{\psr}{B0540-69}

\begin{document}

\title{The Thermal X-ray Emitting Shell of LMC Supernova Remnant 0540-69.3}
\author{Una Hwang (1,2), Robert Petre (1), Stephen S. Holt (3), and Andrew
E. Szymkowiak (1)} 
\affil{(1) NASA Goddard Space Flight Center, Greenbelt, MD 20771 \\ 
(2) Department of Astronomy, University of Maryland, College Park, MD 20742\\ 
(3) F. W. Olin College of Engineering, Needham, MA 02492}

\begin{abstract}
We use data from the Advanced CCD Imaging Spectrometer (ACIS) on the
Chandra X-ray Observatory to image the shell surrounding the pulsar
\psr, and to measure its spectral properties.  Weak emission line
features, notably of Fe, Mg and Si, indicate that the shell is
composed primarily of ambient material heated by the blast wave.  The
shell emission in the east is faint and circular in appearance, with a
temperature and ionization age of approximately 5 keV and $10^{10}\
\rm{cm}^{-3}$ s, respectively.  The emission in the west is brighter
and has a more complex, distorted morphology, and is fitted with
significantly lower average temperatures and higher ionization ages.
An image with photon energies above 2 keV shows two hard arcs of
emission located diametrically opposite each other at the outer
boundary of the remnant shell; the spectra from these regions could
include a substantial nonthermal component.
\end{abstract}

\section{Introduction}
The 50 ms pulsar \psr\ in the Large Magellanic Cloud (LMC) has often been
compared to the Crab: both are young and bright, have similar periods
and ages, and have been detected at X-ray through radio wavelengths.  The
pulsar spindown age is 1660 yr (Seward, Harnden, \& Helfand 1984),
making 0540-69.3 the second youngest known supernova remnant in the Large
Magellanic Cloud after SNR 1987A, which is located nearby in the same
neighborhood of 30 Doradus.  

In the radio, PSR \psr\ is surrounded by a well-defined shell of
radius $\sim 30''$ (Manchester, Staveley-Smith, \& Kesteven, 1993).
This supernova remnant has the distinction of being classified as
oxygen-rich because it appears in the optical as a small 4$''$ radius
shell of emission bright in [OIII], presumably from ejecta (Mathewson
et al. 1980, Chanan, Helfand, \& Reynolds 1984, Kirshner et al. 1989,
Caraveo et al. 1992); some patchy [OIII] emission is also detected at
a larger radius just south of the indentation in the radio shell to
the west (Seward \& Harnden 1994, Mathewson et al. 1980).  Both the
presence of the pulsar and the oxygen-rich ejecta are consistent with
a massive progenitor for \snr.  Reynolds (1985) constructed a scenario for
this remnant based on all the data then available, pointing out that the
optical emission probably arises from ejecta that have been shocked
and accelerated by the expanding bubble of relativistic gas associated
with the pulsar.  This situation is similar to that in the Crab, where
the weak [OIII] emission surrounding the synchrotron nebula is
produced by a radiative shock driven into the freely-expanding ejecta
by a bubble of relativistic gas, while the bright optical filaments
are formed via the Rayleigh-Taylor instability (Hester et
al. 1996). The probable age of the remnant 0540-69.3 is 800$-$1000 yr,
according to Reynolds (1985); the age of 762 yr implied by the optical
expansion velocities (Kirshner et al. 1989) is likely to be an
underestimate because the ejecta are expected to have been accelerated
by the pulsar nebula.

The pulsar nebula dominates the X-ray emission from \snr\ (Clark et
al. 1982), but diffuse X-rays corresponding roughly to the radio shell
were imaged with the ROSAT High Resolution Imager (Seward \& Harnden,
1994).  A recent Chandra High Resolution Camera calibration
observation provides a much improved image clearly showing a distorted
X-ray emitting shell, much brighter towards the west (Gotthelf \& Wang
2000).  The diffuse X-rays are presumed to be thermal emission
associated with the blast wave, but neither the ROSAT nor Chandra
imaging observations provide spectral information.  Kaaret et
al. (2001) used these and other Chandra calibration observations to
study the pulsar, for which we present our results in a separate paper
(Petre et al. 2001, in preparation).

The recent XMM-Newton observation of \snr\ (van der Heyden et al.
2001) gave the first conclusive proof of thermal X-ray emission
associated with the remnant. Spectra from the Reflection Grating
Spectrometers (RGS) on XMM show blueshifted (radial velocities $\sim
-2400$ km/s) emission lines of O VIII from the southwestern part of
the remnant, where the line emission is most prominent.  A temperature
of $kT$=0.6 keV and an ionization age (the product of electron density
and the time since shock heating) of $n_et$=2.5$\times 10^{10}$
cm$^{-3}$ s are inferred for this region.  The roughly 1$'$ diameter
of the remnant, compared to the spatial resolution of the XMM mirrors,
interferes with the ability to make more definite statements about the
origin of the thermally emitting gas.  The thermal spectra are all
blended with emission from the pulsar nebula, and while the RGS
provide excellent high-resolution soft X-ray spectra, their
sensitivity to X-rays with energies above about 2 keV is low.

We report here on Chandra X-ray observations of \snr\ with the
Advanced CCD Imaging Spectrometer (ACIS).  These data show a complete
ring of X-ray emission surrounding the pulsar with a striking
correspondence to the radio shell.  Furthermore, ACIS allows us to
isolate the diffuse thermal X-ray emission from this remnant from that
of the pulsar, and to detect its spatial variations.

\section{Observations and Images}

Chandra observed \snr\ on 22-23 November 1999 for 27.4 ks with the
primary back-illuminated CCD chip ACIS-S3.  The remnant was imaged
entirely on readout node 1 of the CCD, and the data were
accumulated in faint mode; the reader is referred to the Chandra
documentation for technical details about ACIS and
Chandra\footnote{(http://asc.harvard.edu/udocs/docs/docs.html)}.  The
total count rate was stable during this observation, with no indication
of strong background flares.

Figure 1 shows the ACIS broadband (0.3-10 keV) image, both raw and
adaptively smoothed with a boxcar filter of variable size containing a
minimum of 25 counts per smoothing beam; the 5 GHz radio image and
contours from Manchester et al. (1993) are also shown on roughly the
same angular scale.  The high X-ray count rates in the vicinity of the
pulsar result in two image artifacts: 1) low detected counts where the
true count rate exceeds the pile-up threshold (this effect is not
visible in Figure 1 as the image of the pulsar nebula has been
overexposed to enhance visibility of the faint shell), and 2)
symmetrical trails along the y-axis of the detector (diagonal in the
figure, which is oriented with north towards the top of the page) due
to trapping of large charge amounts in the piled-up region.  Weisskopf
et al. (2000) give a brief but informative discussion of pile-up in ACIS.

The X-ray shell imaged by ACIS closely follows the radio morphology at
5 GHz, as can be seen by comparing the panels of Figure 1.  In the
east, the emission appears circular, with a slight enhancement at the
limb.  The brightest portions of the shell are to the west, and the
shell is markedly indented in the northwest.  The angular distance
from the pulsar to the edge of the shell varies from 20$''$ (5 pc for
a distance of 50 kpc to the LMC) at the northwest indentation, up to
about 40$''$ (10 pc) in the faint portions of the shell to the east
and south.  The bright spot just south of east in the 5GHz radio image
is not comparably bright in X-rays, but the bright radio and X-ray
features are otherwise well matched.
 
X-ray color images of the remnant are shown in Figure 2.  The left
panel shows the ratio image of counts with energies below 1 keV
relative to counts with energies above 1 keV in each 2$''$ pixel.  As
expected, the
%scale 0.5'' binned by 4 -> 2'' (2-20-01)
hardest emission (with ratios as low as 0.06) is associated with the
plerion in the center, but hardness variations are clearly evident
around the shell.  In particular, soft emission is associated with the
bright, high surface brightness emission throughout the west (ratios
of 2$-$4), and to a lesser extent with the region just north of the
plerion.  The eastern part of the remnant is relatively much harder
(ratios of about 0.5).

The right panel of Figure 2 shows the ACIS image for events with pulse
heights corresponding to energies in excess of 2 keV.  Two faint arcs
are located roughly equidistant from the pulsar on opposite sides of
the remnant; the east and west hard arcs contain 30 and 57 counts at
energies above 2 keV, respectively.  Both arcs lie at the outer
boundary of the shell, with the brighter arc in the west being just south
of the indentation.  There are no obvious radio counterparts to these
arcs.

\section{Spectral Analysis}

Approximately 15,000 counts were collected from the SNR in this
observation, excluding the pulsar readout trail and a 7$''$ radius
region in the center enclosing both the pulsar nebula and the
optically emitting oxygen ring.  We carried out spectral fits at
energies between 0.5 and 10 keV for different regions of the shell,
shown in the middle panel of Figure 1.  Our goal was to obtain
measurements of the average temperature, ionization age, and element
abundances for each region.  The region eastward (to the left) of the
readout trail was fitted as a single spectrum, labelled ``Outer
East'', while an annulus fitting inside this region was also fitted
separately and labelled ``Inner East''.  There are 1500$-$2000 counts
in the individual spectra, except for the large eastern regions, which
contain 3000$-$4000 counts each.

The spectral response of the backside illuminated CCD chips is
complicated, largely because of the strong position and temperature
dependence of charge transfer inefficiency in the CCDs, with
differences in the detector gain being most pronounced at the
boundaries between readout nodes.  For this reason, we positioned the
source to fall entirely on an area of the focal plane read out by a
single node.  For spectral fitting, we use response matrices for the
appropriate region of the detector based on calibration data from 2000
May for a CCD temperature of $-$110C, and effective areas from the 1999
September release.

At energies below 0.35 keV and above 7 keV, the instrument background
dominates strongly for this source, while the local sky background
towards the LMC is thermal and spatially varying.  The ACIS image of
the entire chip (not shown in the figures) reveals that \snr\ is
located on what appears to be a large, faint, but coherent ring of
thermal gas featuring prominent emission lines, particularly of Si.
We tried local spectral backgrounds taken from various regions of the
S3 chip and found that our spectral results are not particularly
sensitive to the exact background used.

Our basic spectral model is a plane-parallel shock with a single
electron temperature and a range of ionization ages from zero up to a
fitted maximum value (in XSPEC v11.0, Borkowski et al. 2001).  The
ionization age parameterizes the time-dependent, nonequilibrium
ionization of cold gas that has suddenly been heated to a high
temperature by the passage of a shock wave, and is formally the
product $n_et$ of the ambient electron density $n_e$ and the time $t$
since the gas was shocked; collisional ionization equilibrium
corresponds to $n_et \sim 10^{12}$ cm$^{-3}$ s or higher.  This model
provides a simple, but fairly plausible characterization of the
spectra of relatively small spatial regions.

The presence of emission lines of Mg (near 1.3 keV), Si (near 1.85
keV), and Fe (the L blend near 1 keV) in the spectra indicates that
the X-ray emission from the shell has a significant thermal component
(see Figure 3). The element abundances, if fitted freely, are
characteristic of the overall element abundances in the LMC at roughly
0.3 times the solar value (Russell \& Dopita 1992).  The X-ray spectra
of the shell are therefore consistent with the emission arising
predominantly from shock heated ambient material (note that van der
Heyden 2001 infer a roughly 2$-$3 times enhancement in the Ne and Fe
abundance towards the southwest, but do not give errors).  We
therefore present results in Table 1 and Figure 3 for fits to a single
shock component where the element abundances have been fixed at the
LMC values (He=0.89, C=0.30, N=0.12, Fe=0.36, Russell \& Dopita, 1992;
O=0.19, Ne=0.29, Mg=0.32, Si=0.31, S=0.36, Ca=0.34, Hughes, Hayashi,
\& Koyama, 1998; given by number relative to H relative to the solar
photospheric abundances of Anders \& Grevesse 1989).

The fits give lower temperatures and higher ionization ages for the
regions NW, SW, and S in the western part of the remnant than for
regions in the eastern half.  We did additional fits to allow for an
additional underlying hard spectral component in the west by using
either a thermal shock model with the temperature and ionization age
fixed at their values for the Outer East, or a power-law with a fixed
photon spectral index of 2.7; these results are also shown in Table 1.
The fitted temperatures for NW and S are not affected, but the
temperature for SW and the Hard Arcs, which had been somewhat higher
than in the other western regions for the single component fits, are
reduced significantly.  Ionization ages in all regions tend to be
higher than their values for only one spectral component.  The actual
statistical significance of adding the hard spectral components is
low, but the results demonstrate how the spectral complexity can
affect the interpretation of the fitting results.  The relative
contribution of the hard component is indicated by the fractional
detected flux given in the table.  This is the ratio of the detected
flux in the the hard component, i.e., either the 5 keV shock or the
power-law, relative to the total detected flux at energies 0.5$-$10
keV.  This fractional flux is 3$-$4 times higher in the Hard Arcs than
in any other region.

One puzzle is that the fitted column densities for the spectrally soft
regions in the west are between $2-3 \times 10^{21}$ cm$^{-2}$,
roughly a factor of two lower than in the east.  Because of the
complicated low energy response of ACIS, we are not yet able to
ascertain whether this difference indicates a true variation in the
obscuring column.  If the column density is fixed at $4 \times
10^{21}$ cm $^{-2}$, closer to the value found in the east, the fits
are slightly worse, with the reduced $\chi^2$ increasing by up to 0.3.
The only significant change in the fitted temperatures is the 40\%
change in the SW; the best-fit ionization ages all tend to decrease by
up to a factor of two (see the Table).

\section{Discussion}

Because the X-ray spectrum of the shell of \snr\ is evidently
dominated by emission from ambient material swept up by the forward
shock, we estimate evolutionary parameters assuming that the remnant
shell is in the adiabatic phase.  We fitted a Sedov model to the
spectrum of just the eastern half of the remnant to avoid the
complications in the west.  The fit yields parameters listed at the
bottom of Table 1.  The average radius in the east is 30$''$ measured
from the position of the pulsar, yielding a physical radius of 7 pc
for an assumed distance of 50 kpc to the LMC.  Based on this radius
and the measured temperature ($kT$ = 4.1 keV) and ionization age
($n_et$ = 3.7 $\times 10^{10}$ cm$^{-3}$ s), the energy of the
explosion is 2$\times 10^{51}$ ergs, the ambient H density is 0.8
cm$^{-3}$, and the swept-up mass is 40 M$_\odot$.  The deduced age of
the remnant is 1500 yr, which is consistent with the pulsar spin-down
age $P/(2\dot{P})$ = 1660 yr (Seward, Harnden, \& Helfand 1984).
Seward \& Harnden (1994) deduced similar parameters for the remnant
using its extent in the ROSAT HRI image, having assumed an explosion
energy of 2$\times 10^{51}$ ergs and an age close to the
characteristic pulsar age.

The Sedov age is considerably higher than the age of 800$-$1000 yr
given by Reynolds (1985).  The true age could be lower than suggested
by the Sedov analysis if the remnant is actually still in transition
from the free-expansion phase.  For an age of 1000 year and a radius
of 7 pc, the average expansion velocity must have been about 7000
km/s, implying a mean post-shock temperature of 50 keV that is much
higher than the 5 keV electron temperature actually measured in the
east.  A low electron temperature may be expected if the electrons are
not significantly heated by collisionless processes behind the shock,
and are instead equilibrating slowly by Coulomb collisions with the
ions.  A low level of collisionless heating would be consistent with
the relatively high shock speed, as there appears to be an inverse
correlation between Mach number and the efficiency of collisionless
heating (Laming 1998, Ghavamian et al. 2001).  Another possibility is
that particle acceleration at the shock is nonlinear and highly
efficient (Decourchelle et al. 2000), deflecting energy from the
thermal gas; this could be tested with a proper motion measurement of
the current shock velocity, and deeper spectral observations (see
Hughes et al. 2000).
% but Fermi acceleration appears to be related to the same plasma instabilities that heat the electrons (see Ghavamian, Laming)

A tantalizing suggestion of a possible nonthermal X-ray emission
component in \snr\ is provided by the hard arcs shown in Figure 2.
%Though they do not appear to have radio counterparts
In the best fitting model for the combined arcs (which includes a
power-law component), nearly half of the X-ray flux can be attributed
to a nonthermal component: the corresponding unabsorbed 0.5-10 keV
luminosity in the power-law is several times $10^{34}$ ergs/s (for a
distance to the LMC of 50 kpc), which is comparable to the integrated
nonthermal X-ray luminosity of SN 1006 (Dyer et al. 2001, computed for
a distance of 1.8 kpc to SN 1006 taken from Laming et al. 1996).  A
nonthermal luminosity of this magnitude, together with the bipolar
X-ray symmetry of the arcs, suggests that electrons are being
accelerated to X-ray emitting energies behind the shock in \snr\ in a
manner similar to that in SN1006.

Compared to the XMM-Newton results of van der Heyden et al. (2001), we
obtain a similar temperature $kT$ of 0.6 keV for the regions in the
west, but our fitted ionization ages are generally higher by factors
of a few to ten.  Their fitted ionization age is closer to the typical
value that we measure for the hard emission in the east.  The spectral
parameters measured by the XMM RGS should be robust, as they are
determined by the measured intensities of individually detected O VII
and O VIII emission lines, whereas those determined by Chandra are
driven by the shape of the (relatively weak) Fe L emission; however,
the XMM gratings suffer from the blending of emission from different
regions of the remnant, for which we have shown that there are strong
variations in the spectrum.  The discrepancy in $n_et$ may therefore
stem in part from this blending, and should be investigated with
deeper observations by both instruments.  The blueshifts detected by
XMM in the O VIII lines are difficult to confirm in our data, as the
line emission overall is weak and we can neither directly identify nor
isolate the oxygen lines; indeed, we do not observe strong oxygen
lines even the central region of the remnant that contains both the
plerion and the optical ring (Petre et al. 2001).  The fits to the SW
region are slightly improved if the gains are adjusted to mimic the
Doppler shifts, but fits to the other regions are not improved by the
gain shift; in any case, the fitted parameter values are essentially
unchanged.

Very little of the X-ray emitting gas located between radii of 7 and
30$''$ appears to be enriched with supernova ejecta, since our spectra
are well-fitted using element abundances appropriate for the LMC.  The
ring of optically emitting oxygen ejecta is located very close to the
center of the remnant, but this is not likely to be associated with
the reverse shock, which is not expected to be so close to the remnant
center for some 10,000 yr (Reynolds \& Chevalier 1984).  As mentioned
earlier, the optically emitting ring is believed to be associated with
a shock driven into the inner ejecta by the expanding synchrotron
nebula, as is the case in the Crab (Reynolds 1985).  The reverse shock
should be located relatively close behind the blast wave for a remnant
of 0540-69.3's age, but its presence is not evident in our data.  The
reverse shock might not be well-defined if the ambient medium has a
very low density, but our inferred densities of $\sim$ 1 cm$^{-3}$ are
not particularly low.  Perhaps the remnant has encountered relatively
dense ambient material only relatively recently so that the reverse
shock has not yet had time to develop strongly.  A significant
synchrotron component present throughout the shell could reduce the
apparent line equivalent widths and lead us to infer element
abundances that were too low, but then the good agreement with LMC
abundances would have to be attributed to a strong coincidence.

The western region is clearly more complex than the east.  The
boundary of the shell at the western indentation is sharper and
better-defined than elsewhere in the remnant, and it is the brightest
part of the shell at both X-ray and radio wavelengths.  It is also the
only location on the shell where optical [OIII] emission has been
detected, and XMM-Newton, with its higher spectral resolution and
better sensitivity at low energies, detects X-ray OVIII emission both
from this southwestern region and from the central regions, which
presumably correspond to the optical [OIII] ring (van der Heyden et
al. 2001).  Overall, temperatures in the western region are
significantly lower than in the east, and ionization ages
correspondingly higher (in the SW region, the differences are less
marked, but the hard arcs shown in Figure 2 overlap this region, and
it is possible that they are influencing the spectral fits).  The
emission measures indicate that the densities in the west are
significantly higher than in the east by roughly a factor of ten, and
this is further supported by the higher ionization ages in the west.
An approximate pressure balance is broadly suggested in that regions
with lower temperatures generally have higher ionization ages, but
this does not appear to be borne out by a detailed examination of the
fitted parameters.  We are hesitant to assign too much significance to
the exact numerical values from our fits, however, as they based on
rather simplified models for data that are complex, but limited in
photon statistics.  We instead prefer to state that the correlation of
spectral softness with high surface brightness and the pronounced
indentation are qualitatively consistent with the interaction of the
blast wave with a much denser environment in the west.

The source of denser material in the west is difficult to ascertain,
however.  Cohen et al. (1988) show that a CO cloud is located to the
west of the remnant, but their map has a rather coarse spatial scale
of 8.8$'$, and it is not clear that the cloud is close enough for the
remnant to be interacting with it.  The highest resolution HI map
available (Kim et al. 1999) has a 1$'$ resolution comparable to the
size of the remnant, but does show that the region surrounding \snr\
has a rich HI structure.
%, as does higher resolution
%  ground-based optical H images (R. C. Smith, private communication).

We mention, in closing, that an alternative explanation of the
enhanced radio emission in the western part of the remnant is that
energy is injected by the pulsar in two oppositely directed jets
(Manchester et al. 1993).  Though the radio data show a knot of
emission located opposite the bright indentation in the west, this
knot is not bright in X-rays.  The hard X-ray arcs might be
conjectured to be associated with some process involving jets from the
pulsar, but they probably cannot explain the brightness enhancement:
the arcs appear to be localized to a rather small area, and there is
no comparable brightness enhancement associated with the eastern arc.
Moreover, Gotthelf \& Wang (2000) suggest that there is a hint of a
jet-like feature near the pulsar, and that it is oriented
perpendicular to the arcs so that the overall X-ray morphology of the
pulsar nebula is similar to that of the Crab nebula.

In summary, our spatially resolved observations with the Chandra
Observatory have revealed strong variations in temperature and
ionization age around the X-ray emitting shell of \snr.  These
variations appear to be related to the interaction of the forward
shock with its nonuniform surroundings.  The element abundances are
consistent with the mean for the LMC, indicating that we are primarily
observing interstellar material heated by the forward shock.  The
question of why we have not yet observed the emission from the
reverse-shocked ejecta remains to be answered.  Spectral imaging at
energies above 2 keV have revealed two arcs of emission that are
suggestive of nonthermal emission from shock acceleration.  Their true
nature remains veiled, however, until they can be studied further with
deeper observations.

\acknowledgments 

We are grateful to those who have dedicated their time and effort to
making the Chandra X-ray Observatory a successful mission, and thank
the referee for comments that helped to clarify this paper.

\vfil\eject

\begin{deluxetable}{lllllc}
\tablewidth{0pt}
\tablecaption{Spectral Fits$^{\rm a}$}
\tabletypesize{\tiny}
\tablehead{
\colhead{Region}           & 
\colhead{$\chi^2, \chi^2/\nu$} & \colhead{$N_H$ ($10^{21}$ cm$^{-2}$)}  & 
\colhead{$kT$ (keV)}       & \colhead{$n_et$ (10$^{10}$ cm$^{-3}$ s)$^{\rm b}$} & \colhead{Fractional Hard Flux$^{\rm c}$} }
\startdata
\cutinhead{NEI Single-Temperature Plane Parallel Shock}
Outer East & 179.2, 1.6 & 0.50 (0.47$-$0.52) & 4.8 (3.8$-$6.3) & 1.5 (1.3$-$1.7) & ... \\
Inner East    & 171.2, 1.7 & 0.39 (0.37$-$0.41) & 5.5 (4.4$-$7.0) & 1.9 (1.7$-$2.2) & ... \\
Hard Arcs & 73.6, 1.25 & 0.41 (0.36$-$0.43) & 2.1 (1.6$-$3.0) & 1.2 (1.0$-$1.5) & ... \\
NW   & 103.7, 2.1 & 0.33 (0.30$-$0.40) & 0.64 (0.59$-$0.78) & 28 (11$-$55) & ... \\
S    & 72.6, 1.25 & 0.24 (0.18$-$0.28) & 0.62 (0.59$-$0.66) & 39 (27$-$64) & ... \\
SW   & 114.5, 1.7 & 0.26 (0.23$-$0.30) & 1.33 (1.12$-$1.57) & 3.6 (2.8$-$4.9) & ... \\
NW   & 107.6, 2.2 & 0.50 (fixed) & 0.77 (0.64$-$0.88) & 3.6 (2.4$-$8.1) & ... \\
S    & 108.9, 1.8 & 0.50 (fixed) & 0.47 (0.42$-$0.52) & 13 (7.1$-$21) & ... \\
SW   & 167.3, 2.5 & 0.50 (fixed) & 0.60 (0.55$-$0.65) & 1.9 (1.6$-$2.4) & ... \\
NW   & 104.8, 2.1 & 0.40 (fixed) & 0.65 (0.58$-$0.84) & 19 (7.3$-$33) & ... \\
S    & 92.3, 1.6  & 0.40 (fixed) & 0.55 (0.50$-$0.60) & 18 (11$-$27) & ... \\
SW   & 138.0, 2.0 & 0.40 (fixed) & 0.89 (0.81$-$1.0) & 2.1 (1.8$-$2.7) & ... \\
\cutinhead{Two Shock Components: Fixed component with $kT$=5.0 keV, $nt=1.5\times 10^{10}$ cm$^{-3}$ s}
Hard Arcs & 69.3, 1.2 & 0.43 (0.39$-$0.48) & 1.05 (0.63-1.80) & 1.14 (0.94$-$1.67) & 0.45\\
NW   & 80.2, 1.7 & 0.29 (0.24$-$0.36) & 0.60 (0.55$-$0.64) & 97 ($>$ 40) &  0.26\\
S    & 64.5, 1.1 & 0.21 (0.18$-$0.26) & 0.60 (0.57$-$0.63) & 81 (41$-$212) &  0.18\\
SW   & 108.9, 1.65 & 0.25 (0.21$-$0.28) & 0.73 (0.68$-$1.03) & 13 (5.8$-$26) &  0.32\\
NW   & 85.7, 1.75 & 0.40 (fixed) & 0.54 (0.49$-$0.61) & 37 (22$-$67) & 0.19\\
S    & 89.3, 1.5  & 0.40 (fixed) & 0.50 (0.46$-$0.54) & 23 (14$-$39) & 0.09\\
SW   & 125.0, 1.9 & 0.40 (fixed) & 0.56 (0.45$-$0.69) & 4.4 (3.0$-$8.5) & 0.20\\
\cutinhead{Shock + Power-Law with $\Gamma$ = 2.7}
Hard Arcs & 64.7, 1.1 & 0.46 (0.38$-$0.50) & 0.42 (0.35$-$0.57) & 7.3 (3.2$-$19) & 0.40 \\
NW   & 77.4, 1.6  & 0.28 (0.24$-$0.33) & 0.60 (0.55$-$0.64) & 52 (21$-$112)  & 0.16 \\
S    & 70.9, 1.2  & 0.23 (0.19$-$0.26) & 0.62 (0.58$-$0.63) & 42 (28$-$66) & 0.04 \\
SW   & 113.1, 1.7 & 0.27 (0.24$-$0.32) & 0.88 (0.67$-$1.49) & 6.1(3.2$-$11) & 0.10\\
\cutinhead{Sedov}
Outer East   & 177.4, 1.6 & 0.50 (0.47$-$0.52) & 4.1 (3.4$-$5.7) & 3.7 (3.3$-$4.4) & ... \\
\enddata
\tablenotetext{a}{In all the fits, element abundances have been fixed
at appropriate values for the LMC as described in the text.  The error
ranges listed correspond to 90\% confidence for a single parameter
($\Delta\chi^2$=2.71).}  
\tablenotetext{b}{The ionization age listed in the table for shock
models is the maximum value; the model include a range of values, where the 
minimum has been fixed at zero.}
\tablenotetext{c}{The ratio of detected flux in the second, hard spectral 
component relative to the total detected flux for energies between 0.5$-$10 
keV. The hard component is either the thermal shock component with 
a fixed temperature of 5 keV, or the power-law component with a
fixed spectral index of 2.7.}
\end{deluxetable}

\begin{figure}
\centerline{\includegraphics[scale=0.40]{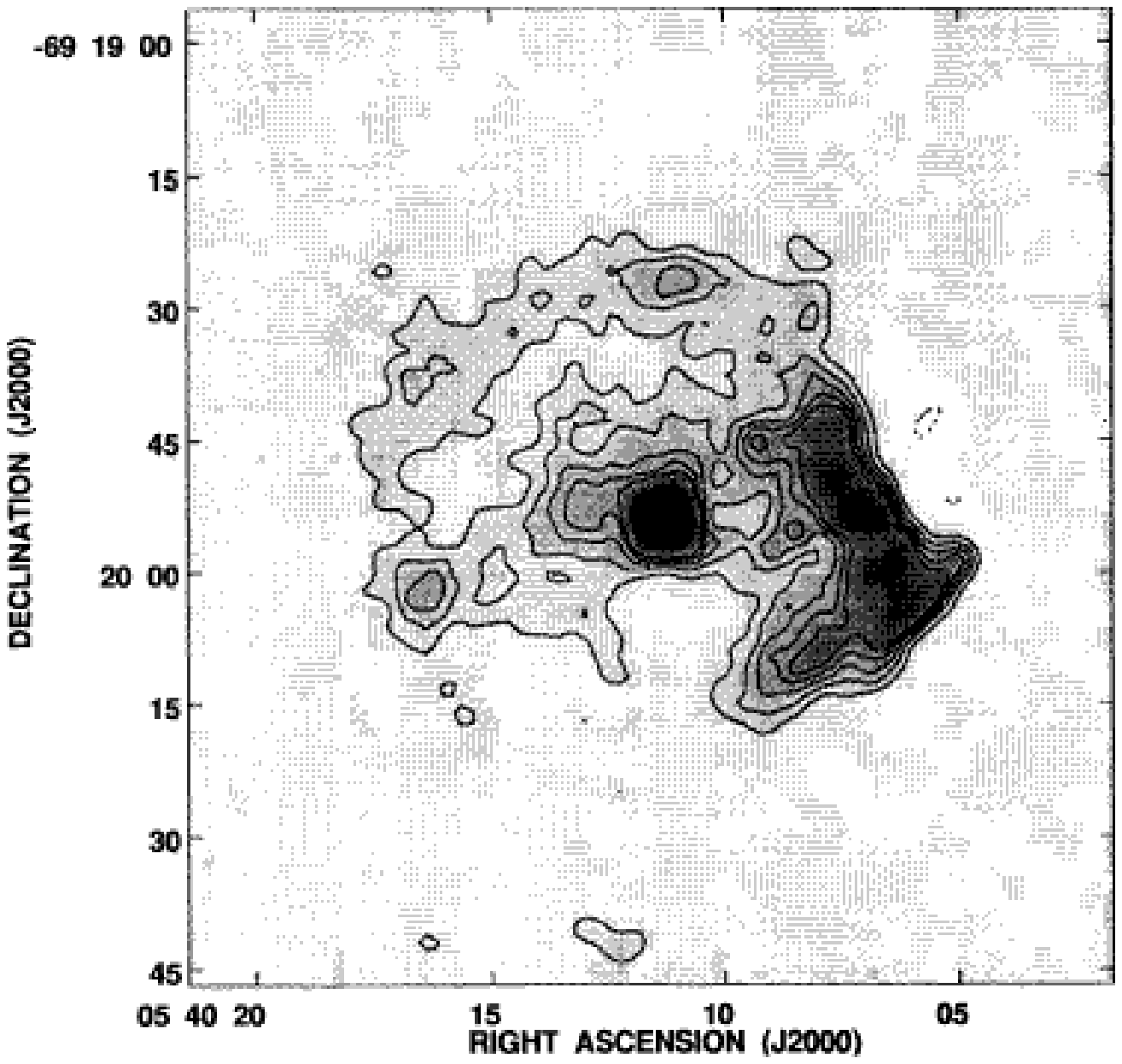}\includegraphics[scale=0.33]{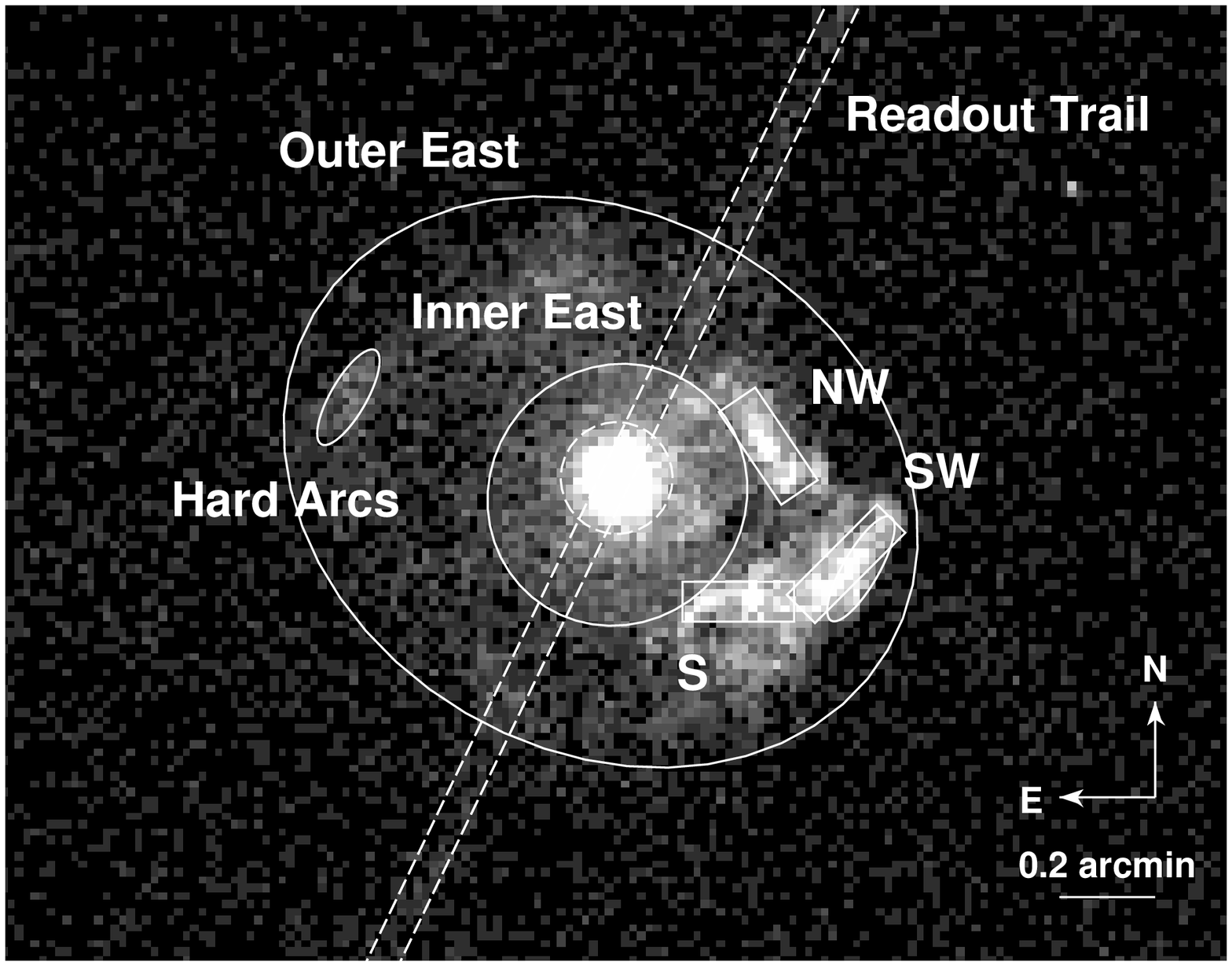}\includegraphics[scale=0.33]{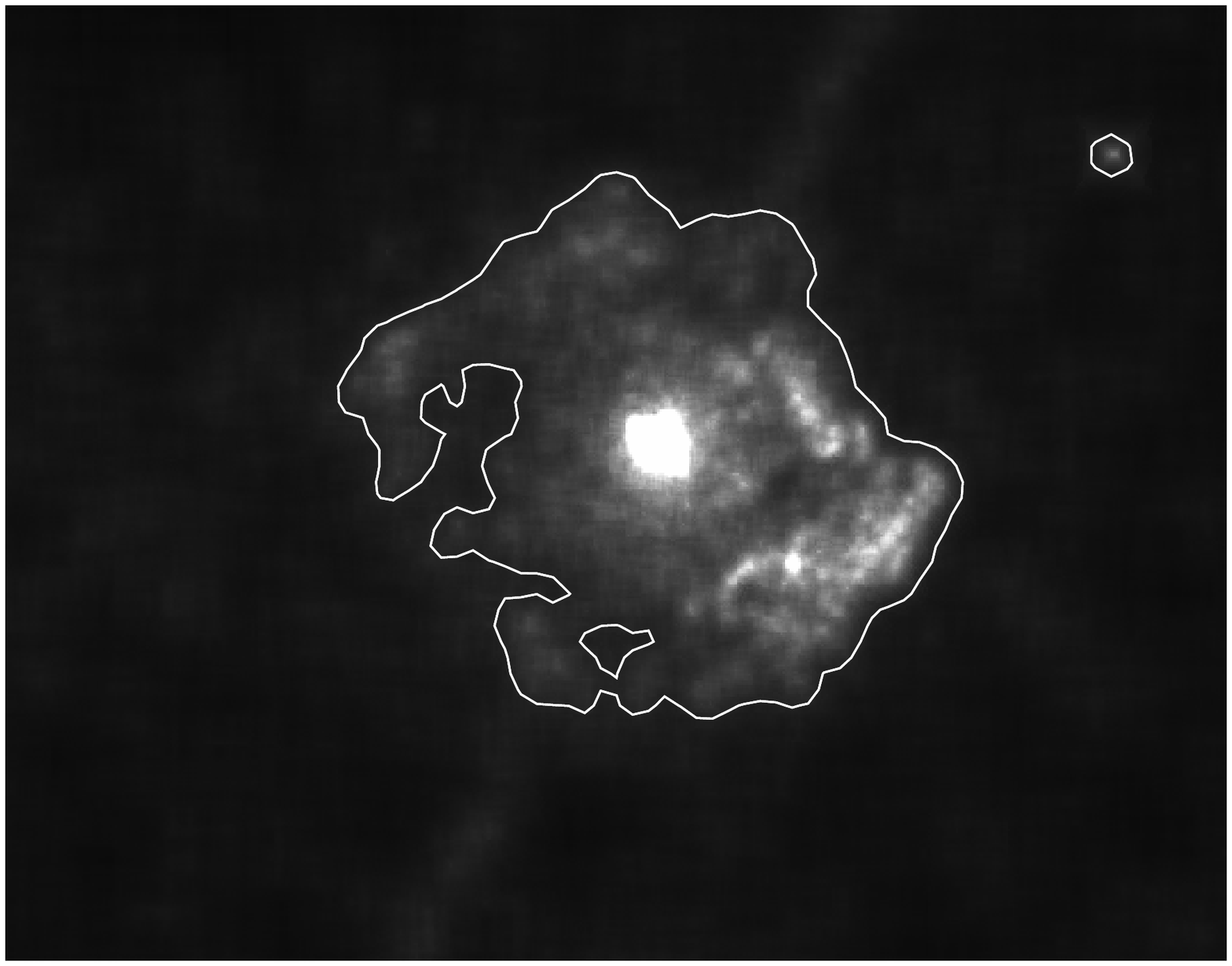}}
\caption{Left: Radio image and contours at 5 GHz taken from Figure~1 of
Manchester et al. (1993).
Middle: On approximately the same angular scale, the unsmoothed
broadband ACIS S3 image of \snr\ in 1$''$ pixels.  The (square-root)
intensity scale has been cut off at 30 counts per pixel to optimize
the visibility of the faint ring.  The faint linear trails leading
diagonally away from the pulsar are artifacts due to charge pile-up in
the detector.  The regions used for spectral analysis are indicated;
the location of the hard arcs in the east and west (see section 2) are
indicated by the two small ellipses.  Right: Broadband ACIS image,
adaptively smoothed with a boxcar filter of varying length containing
a minimum of 25 counts per beam.  The square-root intensity scale has
been cut off at 4 counts per 0.5$''$ pixel.  A single constant
intensity contour corresponding to 0.08 counts per pixel is shown for
comparison with Figure 2a.
% cut off at 6 cnts per 0.25''? pixel after smoothing 2-15-00
}
\end{figure} 

\begin{figure}
%\epsscale{0.8}
\centerline{\includegraphics[scale=0.33]{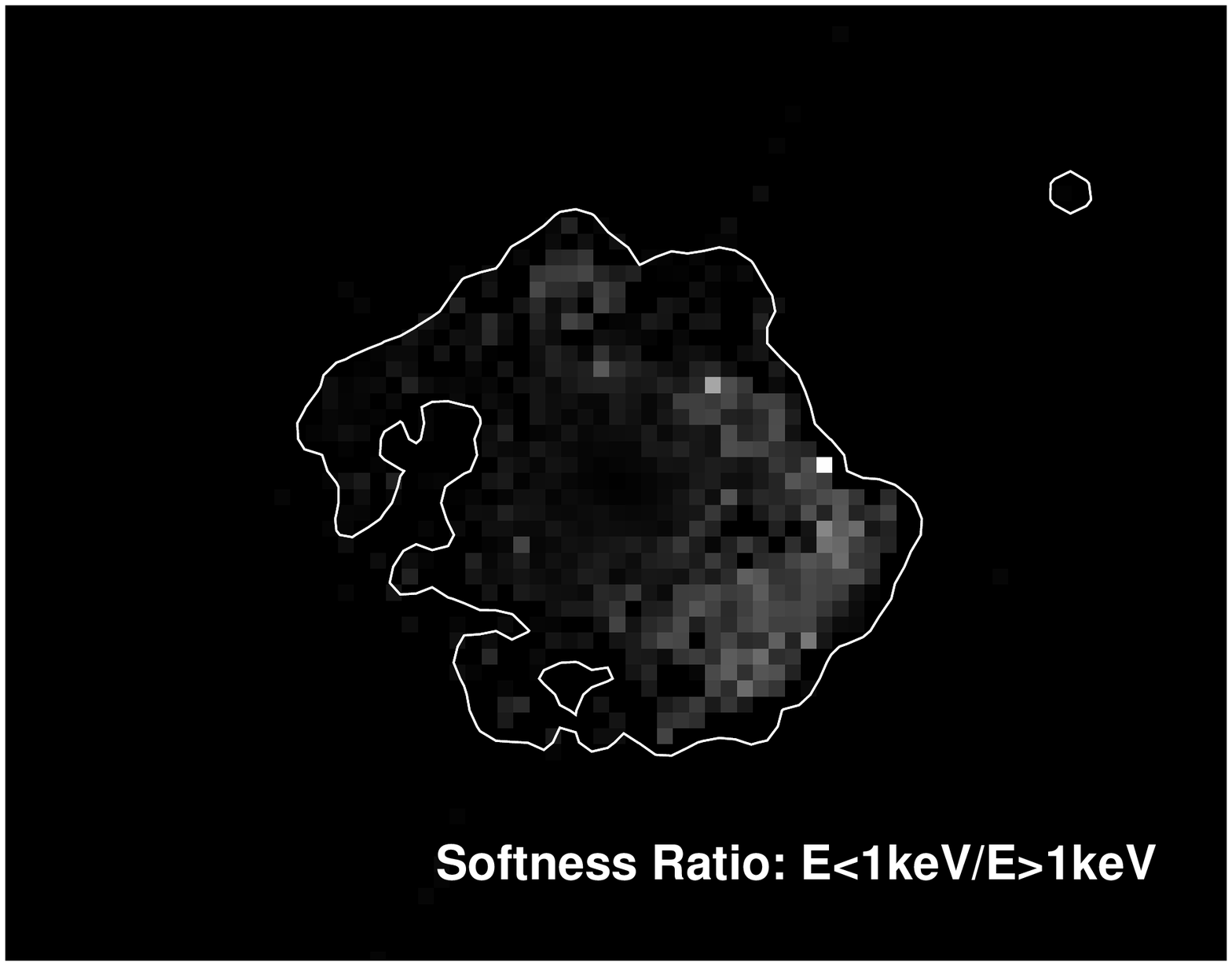}\includegraphics[scale=0.33]{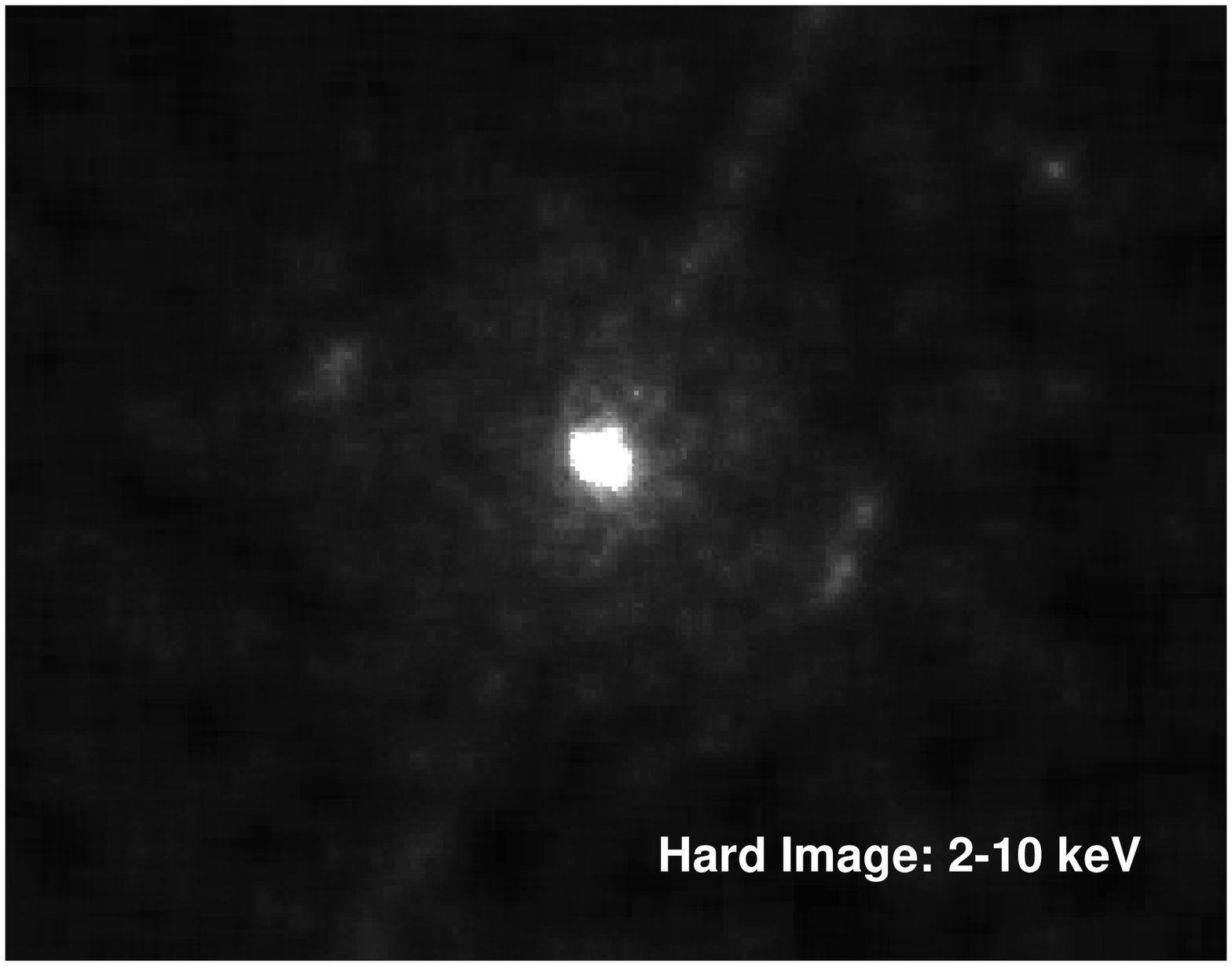}}
\figcaption{Left: ACIS softness ratio map of \snr\ showing the number
of soft counts with energies below 1 keV divided by the number of hard
counts with energies above 1 keV in each 2$''$ pixel.  The same field
of view is shown as in Figure 1, with a linear intensity scale where
white indicates the maximum softness ratio 9.5.  The pulsar nebula at
the center is very hard (with ratios as low as 0.06), as expected, and
appears black.  The superimposed contour is that shown in Figure 1c.
Right: ACIS image of \snr\ for energies between 2$-$10 keV.  The east
and west arcs in this image include roughly 30 and 57 counts,
respectively.  The square-root intensity scale has been cut off above
8 cnts per pixel to enhance the visibility of the arcs.  As in Figure
1, the diagonal lines leading away from the center are artifacts.}
\end{figure} 

\vfil\eject

\begin{figure}
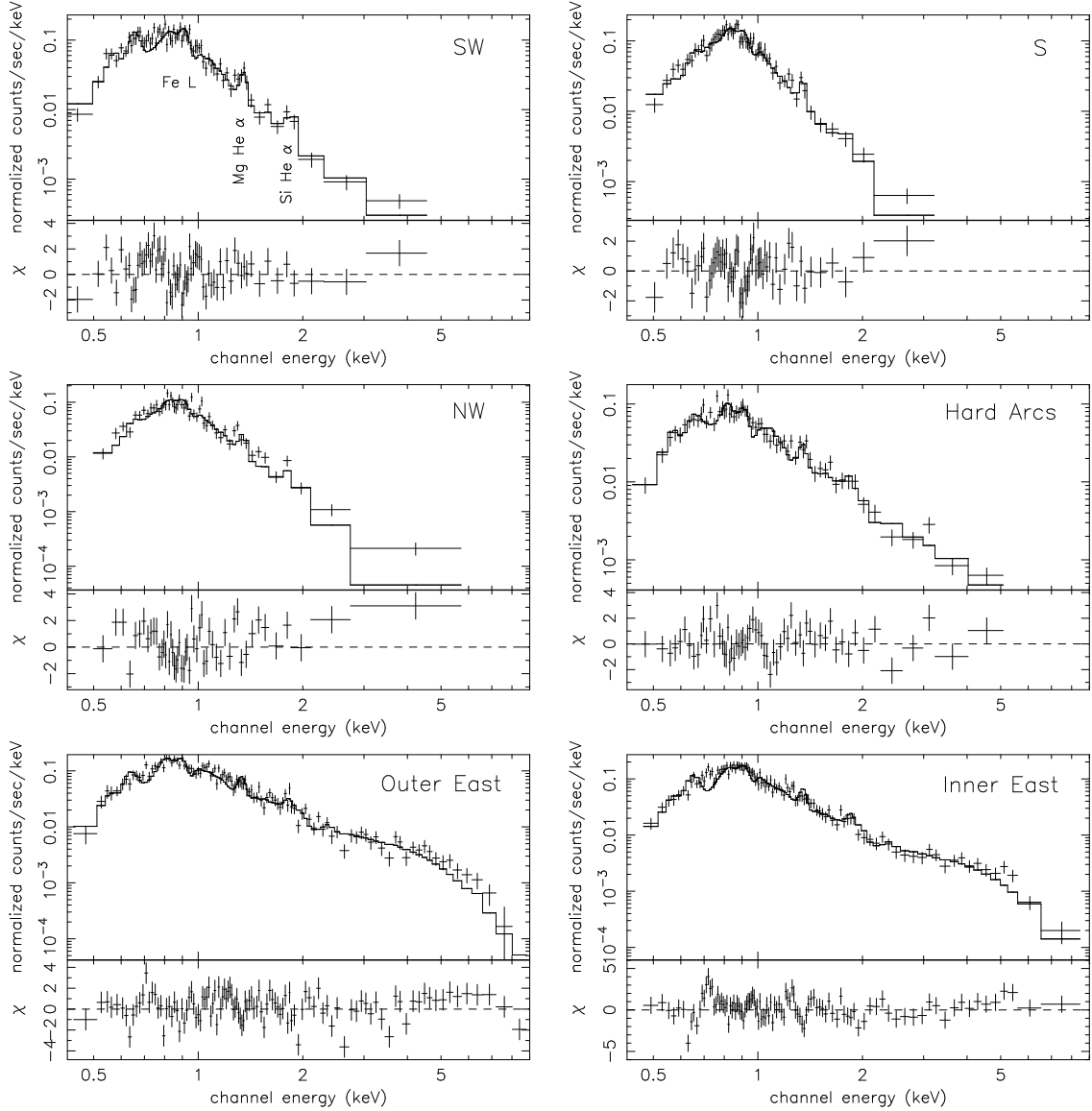

\centerline{\includegraphics[scale=0.30,angle=-90]{fig3a.ps}\hspace{0.2in}\includegraphics[scale=0.30,angle=-90]{fig3b.ps}}
\centerline{\includegraphics[scale=0.30,angle=-90]{fig3c.ps}\hspace{0.2in}\includegraphics[scale=0.30,angle=-90]{fig3d.ps}}
\centerline{\includegraphics[scale=0.30,angle=-90]{fig3e.ps}\hspace{0.2in}\includegraphics[scale=0.30,angle=-90]{fig3f.ps}}
\figcaption{Data and best-fit single temperature plane-parallel shock
model, with element abundances fixed to appropriate values for the
LMC.  The model has been folded through the detector response.}
\end{figure} 

\end{document}